\documentclass[twocolumn,showpacs,superscriptaddress,amsmath,amssymb]{revtex4}

\usepackage{graphicx,stmaryrd,dcolumn}

\begin{document}

\title{Disentangling surface and bulk photoemission using circularly polarized light  }

\author{V.~B.~Zabolotnyy}
\affiliation{Institute for Solid State Research, IFW-Dresden,
P.O.Box 270116, D-01171 Dresden, Germany}

\author{S.~V.~Borisenko}
\affiliation{Institute for Solid State Research, IFW-Dresden,
P.O.Box 270116, D-01171 Dresden, Germany}

\author{A.~A.~Kordyuk}
\affiliation{Institute for Solid State Research, IFW-Dresden,
P.O.Box 270116, D-01171 Dresden, Germany} \affiliation{Institute
of Metal Physics of National Academy of Sciences of Ukraine, 03142
Kyiv, Ukraine}

\author{D.~S.~Inosov}
\author{A.~Koitzsch}
\author{J.~Geck}
\author{J.~Fink}
\author{M.~Knupfer}
\author{B.~B\"{u}chner}
\author{S.-L.~Drechsler}
\affiliation{Institute for Solid State Research, IFW-Dresden, P.O.Box 270116, D-01171 Dresden,
Germany}

\author{V.~Hinkov}
\author{B.~Keimer}
\address{Max-Planck Institut f\"ur Festk\"orperforschung, D-70569 Stuttgart, Germany}

\author{L.~Patthey}
\address{Swiss Light Source, Paul Scherrer Institut, CH-5234 Villigen, Switzerland}


\begin{abstract}
We show that in the angle resolved photoemission spectroscopy (ARPES) near-surface induced fields
can be useful for disentangling the surface and bulk related emission. The jump of the dielectric
function at the interface results in a nonzero term $\operatorname{div}\!\textbf{A}$ in the
photoemission matrix element. The term happens to be significant approximately within the first
unit cell and leads to the circular dichroism for the states localized therein. As an example we
use ARPES spectra of an YBa$_2$Cu$_3$O$_{7-\delta}$ crystal to distinguish between the overdoped
surface related component and its bulk counterparts.
\end{abstract}

\pacs{74.25.Jb, 74.72.Hs, 79.60.-i}

\preprint{\textit{xxx}}

\maketitle

Angle resolved photoelectron spectroscopy is subjected to a known limitation, which is the surface
sensitivity caused by a rather small escape depth of emitted photoelectrons. According to the
universal curve \cite{Seah} the mean free path for the photo electrons in the energy range of
20--70~eV is about 5--10~\AA, which is comparable to the unit cell of the most of the studied
materials. In numerous photoemission studies it is frequently assumed that the spectra from the
first unit cell already reflect the bulk properties of the material and that the surface related
effects are of minor influence. However this is not always the case. ARPES spectra of the high
temperature superconductor YB$_2$Cu$_3$O$_{7-\delta}$ are just one of the examples
\cite{Zabolotnyy, Nakayama}. The signal picked up from the near surface region corresponds to an
unusually high hole doping level and displays no superconductivity, while in the bulk the samples
are characterized by a narrow superconducting transition and uniform doping level, which clearly
shows that the bulk and surface have different properties. Possible way to enhance the bulk
contribution consist in the use of comparatively small (5--10~eV)\cite{Koralek} or large
($\sim$1000 eV)\cite{Suga} excitation energies. However, both of these approaches have certain
restrictions. In the first case it becomes impossible to probe the states in the whole Brilloine
zone, as $k_{\parallel \textup{max}}=\sqrt{2mE_\textup{kin}}/\hbar$ becomes too small, while in the
high energy case significant deterioration of the momentum resolution takes place, not to mention
that both approaches require a specialized light source, which might not always be available.

The fact that the electron escape depth equals $\lambda$ only means that the electron intensity is
attenuated by the factor $\exp{(-z/\lambda)}$, i.e. there are still photoelectrons leaving the
solid from the depth $z >\lambda$, but their intensity is decreasing according to the exponential
law. Therefore the spectra do contain a signal reflecting the bulk properties, but the problem is
how to extract it from under the bright surface contribution. Fortunately the bulk and surface
photoemission differ also due to the pattern of the electromagnetic field that excites the
electrons and this can be used as a ``marker" that makes these two spectral components discernable.
Already in studies of Cu surface states it was shown that the photoemission matrix elements are
strongly modified by the term $\operatorname{div}\!\textbf{A}$, which becomes important at the near
surface region due to the mismatch in the dielectric constant between the solid and the vacuum, and
the calculated dependence of the matrix element on the incidence angle of the linearly polarized
exciting radiation was found to be in a good agreement with experimental data \cite{Pforte}. The
depth down to which the term $\operatorname{div}\!\textbf{A}$ modifies the matrix elements depends
on the field pattern near the surface, exact calculation of which turns out to be a very difficult
task. Relatively simple jellium model employed to describe a free electron metal \cite{Feibelman}
shows that the vector potential of the electromagnetic wave $\textbf{A}(\textbf{r})$ experiences
quickly decaying Friedel-like oscillations with a characteristic scale of a few angstroms. In a
more recent study of TiS$_2$, which is closer to the case we are going to concentrate later on, the
layered structure of the crystal was taken into account \cite{Samuelsen}. From these calculations
it follows that the characteristic thickness $d_\textup{surf}$ of the surface layer is about one
unit cell along the normal to the surface, since in deeper regions $\textbf{A}(\textbf{r})$
practically stops oscillating reaching its bulk value.

\begin{figure*}[floatfix]
\includegraphics[width=\textwidth]{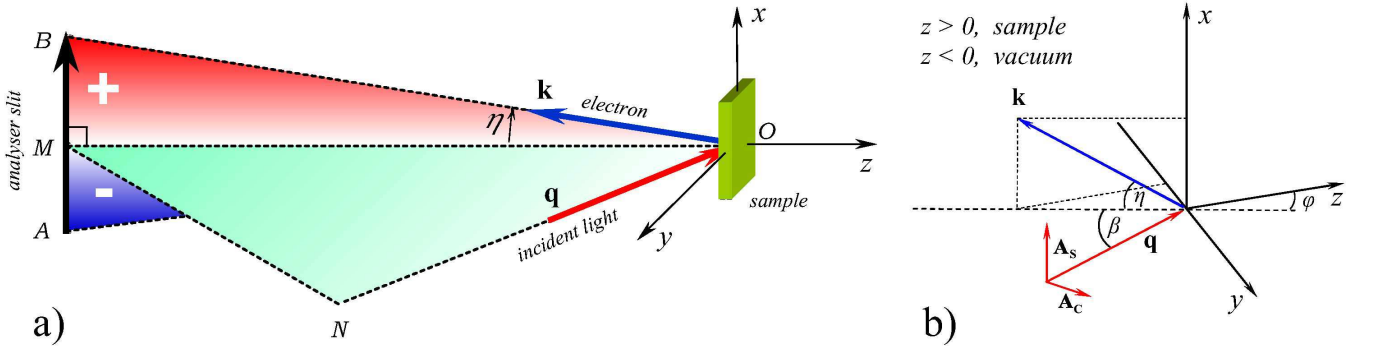}\\
\caption{\label{Images}(color online). Experimental geometry. (a) linear dependence of the circular
dichroism in photoelectron intensity $D=I^\textup{cr}-I^\textup{cl}$ on the angle  $\eta$. The
red-white-blue color scale denotes the strength of the dichroism.
(b) decomposition of a circularly polarized wave in two linearly polarized ones with polarization
vectors $\textbf{A}_\textup{S}$ and $\textbf{A}_\textup{C}$ with a phase shift of $\pi/4$.}
\end{figure*}

In this manuscript we show that the discussed term $\operatorname{div}\!\textbf{A}(\textbf{r})$
leads to a circular dichroism for the photoemission from the surface layer and how the dichroism
can be used to obtain information about the states in the bulk and at the surface.

In general, the photoemission matrix elements depend on experimental geometry (i.e., mutual
position of the sample, polarization, wave vector of incident light $\textbf{q}$ and the direction
of emitted photoelectron $\textbf{k}$). In Fig~1. we depict the geometry of our experiment. The
entrance slit of the energy analyzer is vertical and its position is fixed in space, so that the
ARPES spectrum, the so called energy-momentum distribution, is just a 2D distribution of
photo-intensity as a function of energy and the angle $\eta$ at which electrons enter the analyzer
entrance slit. For a fixed sample position the wave vector $\textbf{k}$ of the photoelectron and
its projections $k_x$ and $k_y$ on the sample surface are uniquely defined by the angle $\eta$ and
can be easily estimated from the drawing. The incident beam lies in the horizontal plane $MON$ with
${\angle}{MON}\equiv\beta=45^{\circ}$. Different parts of reciprocal space can be probed via
rotation of the sample around the pivot point $O$. For instance in Fig.~1b we show the sample (the
crystal primary axes) after rotation around the $x$-axis by the angle $\varphi$, when the
projection $k_y$ becomes negative in contrast to Fig.~1a, where $\varphi=0$ and $k_y=0$.

The photoemission matrix element can be estimated as a probability of a
transition between the initial state $|i\rangle$ and the final state
$|f\rangle$, which is given by the Fermi golden rule \cite{Landau}:
\begin{equation}
 w_{\textup{i}\shortrightarrow\textup{f}} \sim \frac{2\pi}{\hbar}{\bigl|\langle f| \hat{H}_\textup{pert}| i \rangle\bigr|}^2
 \delta(E_\textup{f} - E_\textup{i} - \hbar\omega) \textup{,}
\end{equation}
where the perturbation to the system Hamiltonian
$\hat{H}_\textup{pert}=-\frac{ei\hbar}{mc}(\textbf{A}\nabla+\frac{1}{2}\operatorname{div}\!\textbf{A})$.
The incident circularly  polarized wave $\textbf{A}(\textbf{r})$ can be represented as a sum of two
plane waves (Fig.~1~b):
\begin{equation}
\textbf{A}(\textbf{r})=\textbf{A}_\textup{C}\cos(\omega t - \textbf{qr})
+\textbf{A}_\textup{S}\sin(\omega t - \textbf{qr}).
\end{equation}
It is easy to show that for such a periodic perturbation
\begin{equation}
\begin{split}
 w_{\textup{i}\shortrightarrow\textup{f}} \sim \frac{2\pi}{\hbar}{\bigl|\langle f| \hat{V}_\textup{C} - i \hat{V}_\textup{S} | i \rangle\bigr|}^2
 \textup{, where}  \vspace{0.4em}       \\
\hat{V}_\textup{C,S}=-\frac{ei\hbar}{mc}(\textbf{A}_\textup{C,S}\nabla+\frac{1}{2}\operatorname{div}\!\textbf{A}_\textup{C,S})
\textup{.}
\end{split}
\end{equation}
\begin{figure*}[!t]
\includegraphics[width=16cm]{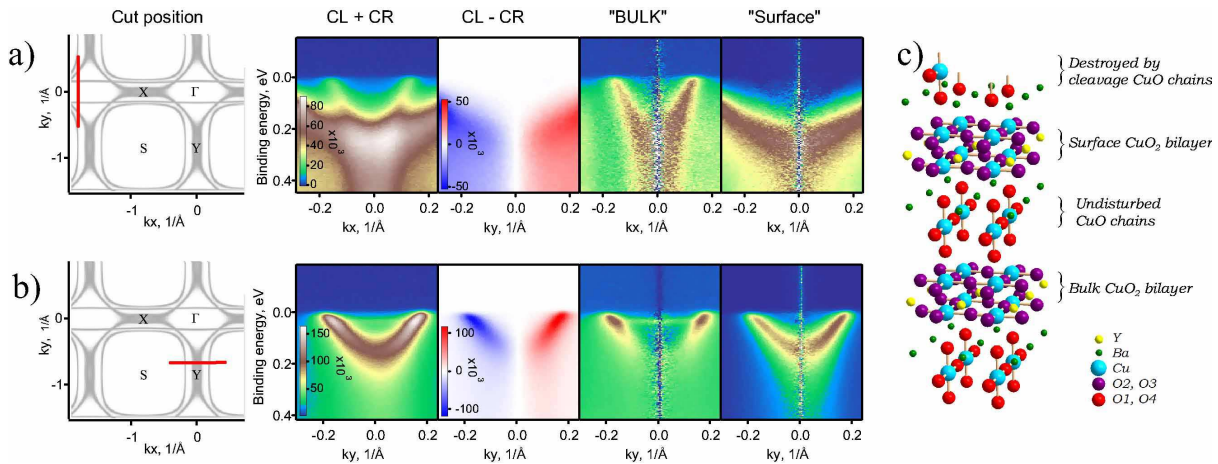}\\
\caption{\label{Images}(color online). (a, b) panels from left to right: position of the
energy-momentum cuts in the reciprocal space; sum of the spectra measured with opposite circular
polarizations; their difference; extracted bulk component; extracted surface component. Color scale
bars show the absolute intensity. In the block (a) the process of disentangling of the surface
related anibonding band and the chain band is demonstrated ($h\nu=50\;$eV). Block (b) depicts the
same procedure, but for the  surface and bulk components of the bonding band ($h\nu=55\;$eV). The
antibonding band in this case is suppressed by the unfavorable matrix elements. (c) Structure of
YBa$_{2}$Cu$_{3}$O$_{7-\delta}$ near the cleavage plane.}
\end{figure*}

    At this stage the particular pattern of the near surface induced field comes into play. Macroscopically,
from the continuity condition for the components of \textbf{D} and
\textbf{E} vectors one has:
\begin{equation}
\begin{array}{rcl}
 \textbf{E}_{\parallel}^\textup{vacuum}  & = &  \textbf{E}_{\parallel}^\textup{sample} \textup{,} \vspace{0.4em}      \\
 \textbf{E}_{\perp}^\textup{vacuum} = \textbf{D}_{\perp}^\textup{vacuum}  & =  & \textbf{D}_{\perp}^\textup{sample} = \varepsilon \textbf{E}_{\perp}^\textup{sample}\textup{,}       \\
\end{array}
\end{equation}
here subscripts $\parallel$/$\perp$ indicate vector components parallel/perpendicular to the sample
surface, and $\varepsilon$ is the sample permeability. Using the gauge with the scalar potential of
the electromagnetic field taken to be zero, the last equation can be rewritten as:
\begin{equation}
\begin{array}{c}
 \textbf{A}_{\parallel}^\textup{vacuum}    =   \textbf{A}_{\parallel}^\textup{sample} \textup{,~~}
 \vspace{0.4em}
 \textbf{A}_{\perp}^\textup{vacuum}        =   \varepsilon \textbf{A}_{\perp}^\textup{sample}\textup{.}       \\
\end{array}
\end{equation}
Microscopically, as has previously been pointed out, the jump in the vector potential appears as a
smooth transition with a spatial extension $d_\textup{surf}$ of about one lattice parameter $c$,
within which the vector potential \textbf{A}(\textbf{r}) changes from its vacuum value to the bulk
one.  For the discussed geometry $\operatorname{div}\!\textbf{A}_\textup{S}=0$ due to the boundary
conditions, and $\operatorname{div}\!\textbf{A}_\textup{C}$ can be approximated as:
\begin{equation}
\begin{split}
\operatorname{div}\!\textbf{A}_\textup{C}&\approx
\frac{\textbf{A}^\textup{sample}_\perp-\textbf{A}^\textup{vacuum}_\perp}{d_\textup{surf}} =
\frac{(1/\varepsilon-1)\textbf{A}^\textup{vacuum}_\perp}{d_\textup{surf}}  \vspace{0.4em}\\
&=CA^\textup{vacuum}_{z}=C|\textbf{A}_\textup{C}|\sin(\beta-\varphi),
\end{split}
\end{equation}
where $C$ is a complex constant that effectively accounts for the width of the transition layer and
sample permeability. To estimate the matrix element (3) we assume that the final state is a plane
wave: $|f\rangle=|e^{i\textbf{kr}}\rangle$, with \textbf{k} being the quasi-momentum of the exited
photoelectron. Leaving out the irrelevant coefficients yields:
\begin{equation}
\begin{split}
w_{\textup{i}\shortrightarrow\textup{f}} & \sim \Bigl| \bigl\langle i | \textbf{A}_\textup{C}\nabla
+ \frac{1}{2}\operatorname{div}\!\textbf{A}_ \textup{C} - i\textbf{A}_\textup{S}\nabla |
e^{i\textbf{kr}} \bigr\rangle\Bigr|^2 \\
&=\bigl| \bigl\langle i | e^{i\textbf{kr}} \bigr\rangle\bigr|^2\Bigl| \textbf{A}_\textup{C} i
\textbf{k} + \textbf{A}_\textup{S}\textbf{k}+ \frac{1}{2}\operatorname{div}\!\textbf{A}_ \textup{C}
\Bigr|^2.
\end{split}
\end{equation}
To get the matrix element for the opposite circular polarization one just needs to reverse the
direction of the vector $\textbf{A}_\textup{S}$, therefore for the dichroism we obtain:
\begin{equation}
\begin{split}
D \equiv
dw^\textup{cr}_{\textup{i}\shortrightarrow\textup{f}}-dw^\textup{cl}_{\textup{i}\shortrightarrow\textup{f}}
\sim \bigl| \bigl\langle i | e^{i\textbf{kr}} \bigr\rangle\bigr|^2
\Bigl\{\Bigl|\textbf{A}_\textup{C} i \textbf{k} + \textbf{A}_\textup{S}\textbf{k} \\
+\frac{1}{2}\operatorname{div}\!\textbf{A}_ \textup{C} \Bigr|^2
 -
\Bigl|\textbf{A}_\textup{C} i \textbf{k} - \textbf{A}_\textup{S}\textbf{k}
+\frac{1}{2}\operatorname{div}\!\textbf{A}_ \textup{C} \Bigr|^2 \Bigr\}.
\end{split}
\end{equation}
Since $|\textbf{A}_\textup{C} i \textbf{k} - \textbf{A}_\textup{S}\textbf{k}|=
|\textbf{A}_\textup{C} i \textbf{k} + \textbf{A}_\textup{S}\textbf{k}|$ there will be no circular
dichroism if $\operatorname{div}\!\textbf{A}_ \textup{C}$ equals zero. It is worth to mention that
this is a rather general statement true for any geometry as long as the final states can be well
approximated by the plane waves \cite{Dubs}. After simple computations, taking into account the
experimental geometry, the expression (8) reduces to the final form that we are going to use:
\begin{equation}
\begin{split}
D \sim \bigl| \bigl\langle i | e^{i\textbf{kr}}
\bigr\rangle\bigr|^2\operatorname{Re}(C)|\textbf{A}_\textup{C}|
|\textbf{A}_\textup{S}||\textbf{k}|\sin(\eta)\sin(\beta-\varphi).
\end{split}
\end{equation}
An important consequence of this formula is that in the described geometry the photoemission signal
arising from the near surface region would exhibit circular dichroism proportional to
$\sin(\eta)\approx\eta$. It is this particularity that can be used to distinguish the contribution
to the spectrum arising from the near surface layer, where the term with
$\operatorname{div}\textbf{A}$ is important, from the one coming from the deeper regions of the
sample. Representing 2D ARPES spectra as a sum of a ``bulk" and the ``surface" component we can
write:
\begin{equation}
\begin{split}
I^\textup{cl}(\eta,\omega)=I^\textup{bulk}(\eta,\omega)+I^\textup{surf}(\eta,\omega)(1 + \alpha\eta),\\
I^\textup{cr}(\eta,\omega)=I^\textup{bulk}(\eta,\omega)+I^\textup{surf}(\eta,\omega)(1 -\alpha
\eta),
\end{split}
\end{equation}
where $\alpha$ accounts for the dichroism strength. The constituent bulk and surface related
components can easily be obtained:
\begin{equation}
\begin{split}
I^\textup{surf}(\eta,\omega) = &\frac{1}{2\alpha\eta}(I^\textup{cl}(\eta,\omega)- I^\textup{cr}(\eta,\omega)),\\
I^\textup{bulk}(\eta,\omega) =&\frac{1}{2}(I^\textup{cl}(\eta,\omega)+I^\textup{cr}(\eta,\omega))
 - I^\textup{surf}(\eta,\omega).
\end{split}
\end{equation}

Now we proceed to the practical application of the derived formulae using as an example ARPES
spectra of YBa$_{2}$Cu$_{3}$O$_{7-\delta}$ (Fig.~2). The spectra were measured from a freshly
cleaved surfaces at $T=30$ K with SES100 SCIENTA electron energy analyzer at the SIS beam line at
Paul Scherrer Institute. The energy and angular resolution were 15\;meV and $0.2^\circ$,
respectively. The excitation energy is given in the caption to the figure, further experimental
details can be found elsewhere \cite{BorisenkoKinks}.

Similar to many other layered compounds its weak $k_z$ dispersion allows for a relatively simple
exposition of ARPES data.  In this case the spectral function depends only on three parameters, so
its argument can be thought of as a point in the 3D space spanned over one energy and two momentum
axes. Therefore the 2D energy-momentum intensity distributions, already mentioned during the
discussion of experimental geometry, are practically almost plane cuts through this 3D space with
the bright features therein corresponding to the traces of the renormalized electronic bands
\cite{Borisenko}. In the leftmost panels of Fig.~2a,b we show a model of the
YBa$_{2}$Cu$_{3}$O$_{7-\delta}$ Fermi surface map that consists of pairs of square-like contours
around the S points, corresponding to the bonding and antibonding bands, and features parallel to
$k_x$ axis, which are the Fermi level crossings of the chain derived band. The red segments denote
the position in the $k$-space of the momentum-energy distribution given in the next panels, which
are: the sum of spectra obtained with the opposite circular polarizations; the difference of the
spectra; the extracted bulk and surface contributions based on formulae (11). To achieve maximal
subtraction of the surface component, parameter $\alpha$ was increased until negative values
appeared.

From the Fig.~2a it follows that the states of the intense overdoped antibonding band are supposed
to be localized at the near surface region, while the chain band has to be of bulk origin as it
displays no dichroism. To understand this one needs to look in detail at the structure of cleaved
YBa$_{2}$Cu$_{3}$O$_{7-\delta}$ shown in Fig.~2c. According to the tunneling experiments
\cite{Edwards, Pan} this crystal cleaves between BaO and CuO layers, which is schematically shown
in the figure as broken bonds and missing atoms. The remnants of the CuO chains, heavily disrupted
by the cleavage, are unlikely to result in photoemission signal that reminds the bulk dispersion of
this band. Therefore the signal from the nearest to the surface CuO$_2$ bilayer should be the
brightest.  In a view of missing chain structure on top of this bilayer, the change of its hole
doping level is not surprising \cite{Zabolotnyy, Nakayama}. The absence of the dichroism for the
chain states simply means that the nearest to the surface chains are already out of the region
where $\textbf{A}(\textbf{r})$ strongly oscillates resulting  in perceivable
$\operatorname{div}\!\textbf{A}$. This also means that the next CuO$_2$ bilayer, which we expect to
be superconducting as it is surrounded by the CuO chains on its both sides, resides in the region,
where $\operatorname{div}\!\textbf{A}$ becomes negligible and should exhibit no dichroism, similar
to the chain band. Indeed, splitting the spectrum of the bonding band (Fig.~2b) into surface and
bulk contributions we see that the surface component reminds the spectrum of the normal  state
taken above $T_\textup{C}$, exhibiting no unusual renormalization.  The spectrum of the bulk
component looks qualitatively different. The strong band renormalization, which is a known
signature of the superconducting state \cite{Bogdanov, Kaminski, Kim, Abanov, Eschrig}, is clearly
visible in the spectrum, supporting the expectation.

Aiming at true quantitative analysis of the disentangled spectra one needs to be cautious as the
dichroism in photoemission experiments is a rather ubiquitous phenomenon. Unlike the classical
linear magnetic dichroism, which rarely exceeds 1\% \cite{Ritchie}, the strength
$(I^\textup{cr}-I^\textup{cl})/(I^\textup{cr}+I^\textup{cl})$ of the described surface related
dichroism exceeds 60\%, so the former can safely be neglected. However, notable dichroism can also
arise from the bulk states \cite{Matsushita}. While the surface related dichroism vanishes at the
normal light incidence (Eq. 9, $\beta=\varphi$), the bulk one is expected to be still non-vanishing
if the experimental setup (including the sample) possesses definite handedness and the final states
substantially differ from the plane waves. To estimate its strength in cuprates we can refer to our
previous results \cite{BorisenkoCP1}, where the total dichroism for the normal light incidence was
shown to be not more than 6\%. For the arbitrary light incidence \cite{BorisenkoCP2} (when the
surface dichroism is vanishing due to condition $\eta=0$) the total dichroism amounts to
$\sim$18\%, but for the sample orientation discussed here, i.e. when one of the crystal primary
axes lies in the $MON$ plane, its value is zero again, so that the surface related dichroism turns
out to be a dominating one.

To conclude, we have stressed the importance of the near surface induced electromagnetic fields for
the interpretation of photoemission data, and pointed out that those might be a beneficial factor
allowing for distinguishing between the surface and bulk photoemission. In some cases it is also
possible to localize the position of the surface and bulk states more precisely, like in the case
of YBa$_{2}$Cu$_{3}$O$_{7-\delta}$, where the nearest to the surface CuO$_2$ bilayer happens to be
overdoped, while the next bilayer can already be treated as the bulk one. The obtained results can
easily be extended to any other experimental geometry.

\end{document}